 \documentclass[preprint, aps, pre, eqsecnum, amsmath,amssymb,showpacs]{revtex4}

 \usepackage[final]{graphicx}  

 \newcommand{\Nabla}{\mbox{\bf\boldmath $\nabla$}}

\begin{document}

 \title{Spiral and Taylor vortex fronts and pulses in axial through-flow}                                

 \author{A.~Pinter, M.~L\"ucke, and Ch.~Hoffmann}                                   
 \affiliation{Institut f\"ur Theoretische Physik, Universit\"at des Saarlandes,      
 Postfach 151150, \\ D-66041 Saarbr\"ucken, Germany}                              

 \date{\today} 
                                                                   
 \begin{abstract}
The influence of an axial through-flow on the spatiotemporal
growth behavior of different vortex structures in  
the Taylor-Couette system with radius ratio $\eta=0.5$ is determined.  
The Navier Stokes equations (NSE) linearized around the basic 
Couette-Poiseuille flow are solved numerically with a shooting method in a wide
range of through-flow strengths $Re$ and different rates of co- and 
counterrotating cylinders for
toroidally closed vortices with azimuthal wave number $m=0$ and for spiral vortex
flow with $m= \pm 1$. For each of these three different vortex varieties we have
investigated ({\it i}) axially extended vortex structures, ({\it ii}) axially 
localized vortex pulses, and ({\it iii}) vortex fronts.
The complex dispersion relations of the linearized NSE for vortex modes with 
the three different $m$ are evaluated for real axial wave numbers for ({\it i}) 
and over the plane of complex axial wave numbers for ({\it ii}, {\it iii}).
We have also determined the Ginzburg-Landau amplitude equation (GLE) approximation 
in order to analyze its predictions for the vortex stuctures 
({\it ii}, {\it iii}). Critical bifurcation thresholds for 
extended vortex structures are evaluated. The boundaries between 
absolute and convective instability of the basic state for vortex pulses
are determined with a saddle-point analysis of the dispersion relations.
Fit parameters for power-law expansions of the boundaries up to $Re^4$ are 
listed in two tables. Finally, the linearly selected front behavior of growing 
vortex structures is investigated using saddle-point analyses of the dispersion 
relations of NSE and GLE. For the two front intensity 
profiles (increasing in positive or negative axial direction) we have determined
front velocities, axial growth rates, and the wave numbers and frequencies of
the unfolding vortex patterns with azimuthal wave numbers $m=0, \pm 1$, 
respectively.  
 \end{abstract}
 
 \maketitle 
                                                                     


 \section{INTRODUCTION} \label{SEC.INTRO}
The Taylor-Couette system \cite{review} of fluid flow in the annulus between
concentric cylinders with the inner 
and the outer one rotating with different velocities is one of the simplest 
examples of a driven nonlinear dissipative system that shows spontaneous pattern
formation out of an unstructured basic state that is stable at small driving
\cite{CRO-HOH}. 
This basic flow state is stationary and axially and azimuthally homogeneous and
shows only a radial variation across the annular gap. It consists of a
superposition of circular Couette flow (CCF) in azimuthal direction 
and of an annular Poiseuille flow (APF) in axial direction if as in our case 
an axial through-flow is imposed. Axially periodic vortex flow solutions 
bifurcate \cite{CHO-IOO,GOL-STE-SCH-LAN} out of this homogeneous 
basic flow when the rotation rate of the inner cylinder is sufficiently 
high. These primary bifurcation thresholds to periodic vortex stuctures have
been the aim of many linear stability analyses of the basic flow state  
\cite{DIP-PRI,TAK-JAN,NG-TUR,LAN-TAG-KOS-SWI-GOL,GEB-GRO,MES-MAR}.

For the radius ratio $\eta =0.5$ and the parameter ranges of rotation
rates and through-flow investigated here in this work three spatiotemporally
differing primary vortex structures are relevant: Rotationally 
symmetric, toroidally closed vortices with
azimuthal wave number $m=0$ that move in downstream direction with the APF --- 
for shortness we call this flow state Taylor vortex flow (TVF) although the 
presence of an axial through-flow modifies the genuine stationary TVF
stucture. And, furthermore, spiral vortex flow (SPI) consisting of either left 
spiral vortices (L-SPI) with $m=1$ or right spiral vortices (R-SPI) with 
$m=-1$.

L-SPI and R-SPI are axial mirror images of each other in the absence of axial 
through-flow with the latter breaking the mirror symmetry of the former. 
While rotating azimuthally into the same direction as the inner cylinder 
L-SPI propagate axially opposite to R-SPI. This spiral dynamics is largely 
induced by the advective properties of the basic flow state.
Furthermore, without through-flow the symmetry degenerate bifurcation treshold 
for these two symmetry degenerate SPI
solutions is simultaneously also the bifurcation threshold for a vortex flow 
solution called ribbons \cite{CHO-IOO}. This solution consists right at threshold
of a linear superposition
of L-SPI and R-SPI with equal amplitude and it becomes further away from 
threshold a genuine nonlinear vortex flow solution. However, here 
we are dealing only with linear vortex flow fields that may be superimposed 
with arbitrary amplitudes as well as wave numbers and that are evolving 
separately from each other according to the linear field equations. Thus we do 
not need to discuss
ribbons separately from our general investigation of linear vortex modes with
general axial and azimuthal wave numbers. 

In this work we quantitatively determine the influence of an axial through-flow
on the spatiotemporal growth properties of linear perturbations of the basic
flow state with azimuthal wave numbers $m=0$ and $m=\pm 1$, i.e., of toroidally
closed vortices and of spiral vortices, repectively. In each case we investigate
({\it i}) axially extended structures, ({\it ii}) pulses of axially localized 
wave packets of vortices, and ({\it iii}) vortex fronts. 

In Sec.~\ref{SEC.SYSTEM} we describe the system, we
briefly review the linearized Navier-Stokes equations (NSE) for the eigenvalue 
problem describing vortex perturbations, and
we give details of our numerical procedure to solve the eigenvalue problem.
In Sec.~\ref{SEC.EXTENDED_VORTICES}
we discuss the spatiotemporal structure, symmetry properties, and
bifurcation thresholds for onset of axially extended
vortex perturbations of the form $e^{i(kz + m\varphi)}$ with real axial wave 
number $k$ and different azimuthal wave numbers $m$ in the absence and presence 
of an axial through-flow. In Sec.~\ref{SEC.LOCAL-PERTUB}
we consider axially localized wave packets 
consisting of superpositions of vortex eigenmodes of the linear NSE. Here we
determine among others the boundary between convective and absolute instability
of the basic flow against growth of vortices with a particular $m$ by a 
saddle point analysis of the linear complex dispersion relation of the NSE
over the plane of complex axial wave numbers. In addition we also determine 
the Ginzburg-Landau amplitude 
equation (GLE) approximation for the dispersion relation for the sake of
comparison. In Sec.~\ref{FRONTS} we evaluate the spatiotemporal properties of
linearly selected vortex fronts using a saddlepoint analysis of the dispersion 
relation. Also here we compare with GLE results. The final section contains a
summary.

 \section{System} \label{SEC.SYSTEM} 
Here we describe the system and we provide definitions and equations. Then we
briefly review the linearized equations for the eigenvalue problem describing 
vortex perturbations of the basic flow state. Finally
we give details of our numerical procedure to solve the eigenvalue problem.
 \subsection{Setup} \label{SEC.SETUP} 
We consider the flow of an incompressible fluid in the annulus between two
concentric cylinders of inner radius $r_1$ and outer radius $r_2$ with a gap width
$d=r_2-r_1$. The boundary conditions at $r_1$ and $r_2$ are no-slip. The angular
velocity of the inner and outer cylinder is $\Omega_{1}$ and $\Omega_{2}$, 
respectively. The associated Reynolds numbers are 
\begin{equation}
R_{1}=\frac{d}{\nu}r_1\Omega_{1}, \quad
R_{2}=\frac{d}{\nu}r_2\Omega_{2},
\end{equation}
where $\nu$ is the kinematic viscosity. An externally imposed axial through-flow
is measured by the axial Reynolds number
\begin{equation}
Re=\frac{d}{\nu} \langle w \rangle
\end{equation}
where the mean axial velocity $\langle w \rangle$ averaged over the annular 
cross section describes the total through-flow. We use also the relative control 
parameters 
\begin{equation}
\mu = \frac{R_{1}}{R_{1c}(Re)}-1,\quad \epsilon = \frac{R_{1}}{R_{1c}(Re=0)}-1,
\end{equation}
measuring the relative distance of the inner Reynolds number $R_{1}$ from
the critical onset $R_{1c}$ of axially extended spiral vortices or Taylor 
vortices in the presence and in the absence $(Re=0)$ of through-flow, 
respectively \cite{epsT}. In this notation 
\begin{equation} \label{epsilon_c}
\mu_c = 0 \quad \mbox{and} \quad 
\epsilon_c (Re) =  \frac{R_{1c}(Re)}{R_{1c}(Re=0)} -1
\end{equation}
is the critical threshold for onset of the vortex flow in question. The relation 
between $\mu$ and $\epsilon$ is 
\begin{equation}
\mu= \frac{\epsilon-\epsilon_{c}(Re)}{1+\epsilon_c (Re)}.
\end{equation}
With infinitely long cylinders the only relevant parameter
characterizing the geometry is the radius ratio $\eta=r_1/r_2$.

The velocity field $\bf u$ of the fluid is described by the Navier-Stokes 
equations (NSE) for incompressible fluid flow 
\begin{equation} \label{NAVIER2}
\partial_t {\mathbf u}= \Nabla^2 {\bf u} -
 R_{1} ({\bf u}\cdot \Nabla){\bf u} - \Nabla p,\quad  \Nabla \cdot {\bf u} = 0.
\end{equation}
Here and in the following we scale positions by the gap width $d$, the velocity
${\bf u}$ by the velocity $r_1\Omega_{1}$ of the inner cylinder, time $t$ by
the momentum diffusion time $d^{2}/\nu$ across the gap, and the pressure $p$ by 
$\rho r_1 \Omega_{1}\nu/d$ with $\rho$ denoting the constant mass density of the
fluid. Furthermore we decompose the velocity field 
\begin{equation}
{\bf u} = u{\bf e}_r + v{\bf e}_{\varphi} + w{\bf e}_z
\end{equation}
into radial ($u$), azimuthal ($v$), and axial ($w$) components
using cylindrical coordinates $r,\varphi,z$.
 \subsection{Basic flow state} \label{SEC.BASIC_FLOW} 
The basic flow state ${\bf u}_0$ that is realized
in the absolutely stable regime of inner Reynolds numbers $R_1$ below the 
thresholds for onset of Taylor and spiral vortex flow is rotationally symmetric,
axially homogeneous, and constant in time. It consists of a linear superposition 
of circular Couette flow (CCF) in azimuthal direction, ${\bf e}_{\varphi}$, and 
of annular Poiseuille flow (APF) in axial direction, ${\bf e}_z$,
\begin{equation}\label{u_0}
{\bf u}_0 = v_{CCF}(r){\bf e}_{\varphi}+w_{APF}(r){\bf e}_{z}
\end{equation} 
without any radial component. Here
\begin{equation} 
v_{CCF}(r)=Ar+B/r,
\end{equation}
and 
\begin{equation}
w_{APF}(r)=Re\frac{r^{2}+C ln(r)+D}{E},
\end{equation}
with 
\begin{eqnarray}
A&=&-\frac{\eta^{2}-\Omega_{2}/\Omega_{1}}{\eta(1+\eta)},\\
B&=&\frac{\eta(1-\Omega_{2}/\Omega_{1})}{(1-\eta)(1-\eta^{2})},\\
C&=&\frac{1+\eta}{(1-\eta)ln(\eta)},\\
D&=&\frac{(1+\eta)\ln(1-\eta)}{(1-\eta)\ln(\eta)}-\frac{1}{(1-\eta)^{2}},\\
E&=&-\frac{1}{2}R_{1}\frac{1-\eta^{2}+(1+\eta^{2})\ln(\eta)}{(1-\eta)^{2}\ln(\eta)}.
\end{eqnarray} 
 \subsection{Linear eigenvalue problem of vortex perturbations} 
 \label{SEC.LIN-STAB-ANALYSIS}
Let $\psi =(u,v,w,p)$ abbreviate the deviation fields from the basic flow state
(\ref{u_0}). Then the general solution of the NSE linearized in the deviation 
fields can be written as a superposition of modes of the form
 \begin{equation}\label{Ansatz}
\psi(r,\varphi,z,t) = \phi(r)e^{i(kz + m\varphi)}e^{\sigma t}
\end{equation} 
with axial wave number $k=2\pi/\lambda$ and integer azimuthal wave number $m$. 
The complex amplitude functions
\begin{equation}
\phi(r) = \left[ U(r),V(r), W(r),P(r) \right]
\end{equation}
depend on the mode indices $k,m$ and the radial coordinate $r$. The characteristic
exponent $\sigma(k,m)$ is in general complex. It is decomposed here as follows
\begin{equation}
\sigma = \Re \sigma + i\, \Im \sigma = \gamma - i \omega 
\end{equation}
into the growth rate $\gamma$ and the characteristic frequency $\omega$ of the 
$k-m$ mode. Substituting the above solution ansatz into the linearized NSE yields
\begin{eqnarray} 
\label{Linear1}
& & \sigma U = \left( \partial_r^2  + \frac{1}{r} \partial_r 
- \frac{1+m^2}{r^2} - k^2 \right) U + 2 F V \nonumber\\
& & \qquad - \partial_r P -im \left( \frac{2}{r^2} V + F U \right)- ik H U \\
\label{Linear2} 
& & \sigma V = \left( \partial_r^2 + \frac{1}{r}\partial_r 
- \frac{1+m^2}{r^2} - k^2 \right) V + 2GU \nonumber\\
& & \qquad -\frac{im}{r} P + im \left( \frac{2}{r^2} U - F V \right)-ikH V \\
\label{Linear3}
& & \sigma W = \left( \partial_r^2  + \frac{1}{r} \partial_r 
- \frac{m^2}{r^2} - k^2 \right) W -ikP \nonumber\\
& & \qquad -im F W - I U-ikH W \\
\label{Linear4}
& & \quad 0 = \partial_r U + \frac{1}{r} U + \frac{im}{r} V + ikW . 
  \end{eqnarray}
The solution of this eigenvalue problem yields the characteristic exponent 
$\sigma$ and the associated eigenfunctions $\phi(r)$ as functions of $k,m$. 
Here 
 \begin{eqnarray} \label{FGHIKonstanten}
F(r) & = &\frac{R_1}{r} v_{CCF}(r) \quad , \quad G = -R_1 A,\\
H(r) & = & R_1 w_{APF}(r) \quad , \quad I(r) = \partial_r H(r)
\end{eqnarray}
are quantities defining the basic flow state (\ref{u_0}). The latter enters via
the linearized advective term of the NSE.

In order to rewrite (\ref{Linear1} - \ref{Linear4}) into a system of first-order
differential equations -- which is advantageous for numerical reasons -- we
introduce three additional complex amplitude functions 
\cite{LAN-TAG-KOS-SWI-GOL}
\begin{eqnarray}
\label{DefX}
X &=& \partial_r U + \frac{1}{r} U - P \, , \\
Y &=& \partial_r V + \frac{1}{r} V \, , \\
Z &=& \partial_r W \, .  
\end{eqnarray}
Using (\ref{DefX}) and the continuity equation (\ref{Linear4}) one can then
eliminate the pressure in Eq.(\ref{Linear2}) by $P=-X-ikW-\frac{im}{r}V.$
All in all one obtains in this way a system of 6 coupled, first-order 
differential equations
\begin{equation} \label{resultEW}
\partial_r {\bf X} = {\cal L}\ {\bf X},
\end{equation}
for the six variables
\begin{equation} \label{start.vector}
{\mathbf X}=\left( U,V,W,X,Y,Z \right)^{T}
\end{equation}
with 
\begin{equation} \label{Result.Operator}
{\cal L} = \left( 
\begin{array}{cccccc} 
-\frac{1}{r} & - i\frac{m}{r} & -ik & 0 & 0 & 0\\  
0 & -\frac{1}{r} & 0 & 0 & 1 & 0 \\ 
0 & 0 & 0 & 0 & 0 & 1\\ 
L  & 2(i\frac{m}{r^2} - F) & 0 & 0 & 0 & 0 \\ 
- 2(i\frac{m}{r^2} + G) & L + \frac{m^2}{r^2}  & \frac{mk}{r} & -i\frac{m}{r} & 0 & 0 \\ 
 I & \frac{mk}{r} &  L + k^2 & -ik & 0 & -\frac{1}{r} \\
\end{array} \right) 
\end{equation}
where 
\begin{equation} \label{defineL}
L = \sigma + \frac{m^2}{r^2} + k^2 + imF + ikH.
\end{equation}
 \subsection{Numerical procedures}  \label{SEC.NUM-PROC1}
 We have solved the eigenvalue equations (\ref{resultEW}) numerically with a
 standard shooting method subject to the six boundary conditions, 
 \begin{equation} \label{result.boundary}
U=V=W=0 \quad \textrm{at} \quad r_1=\eta/(1-\eta) \,\, 
\textrm{and} \,\, r_2 = 1/(1-\eta) \, ,
\end{equation}
which make the eigenvalue spectrum discrete.
To integrate from $r_1$ to $r_2$ we used a fourth-order Runge-Kutta formula 
\cite{STO-BUR} with two step widths ( $\Delta r = 1/200$ and  1/400,
respectively) for a Richardson
extrapolation. A Newton-Raphson method \cite{PRE-TEU-VET-FLA} was then used to 
find the roots of the complex 
 determinant of the $3 \times 3$ matrix which ensures the vanishing of 
 $U,V,W$ at the outer cylinder. We therefore vary in the Newton-Raphson 
procedure only two of the parameters ($\sigma,\eta, R_{1}, R_{2}, Re, m, k$)
\cite{DIP-PRI} that enter into (\ref{resultEW}) while keeping the others 
fixed. In this way we
determine on the one hand the marginal threshold values of $R_1$ and $\omega$ 
with $\gamma=0$ for which the basic state is marginally stable against the
growth of an extended perturbation with given $m$ and real axial wave number 
$k$ at specified parameters $\eta, R_2, Re$. On the other hand we calculate for
given $m, \eta, R_1, R_2, Re$ the complex eigenvalue $\sigma$ over the 
{\em complex} wave number plane -- including as special case also the real 
$k$-axis. In each case we are interested only in the vortex modes with the
largest growth rates for which the associated amplitude functions $\phi(r)$
display the least radial variation with the fewest number of nodes.

We present here results for the radius ratio $\eta=0.5$ in a range of outer
Reynolds numbers $-150 \leq R_2 \leq 50$. In this parameter regime the vortex 
perturbations with the
largest growth rates have in the absence of through-flow azimuthal wave numbers
of either $m=0$ or $m=\pm 1$ \cite{{LAN-TAG-KOS-SWI-GOL}}. We investigate here 
linear properties of such vortices with $m=0$ and $m=\pm 1$ in a through-flow of
Reynolds numbers $0\leq Re \leq 20$.
 \section{Axially extended vortex structures} 
 \label{SEC.EXTENDED_VORTICES}
Here we discuss the spatiotemporal structure, symmetry properties, and
bifurcation thresholds for onset of axially extended
vortex perturbations with real axial wave number $k$ and different azimuthal 
wave numbers $m$ in the absence and presence of an axial through-flow.
 \subsection{Spatiotemporal structure} \label{SEC.STRUCTURE}
Structure and dynamics of the vortex modes (\ref{Ansatz}) are dominated by the 
fact that their phases are constant on any cylindrical surface,
$r=const$, along lines given by the equation
 \begin{equation} \label{z_phase}
z_0 = - \frac{m}{k}\varphi + \frac{\omega(k,m)}{k}t \, .
\end{equation}
Here the constant phase coming from the amplitude $\phi(r)$ is suppressed.
Thus, on the $\varphi-z$ plane of such an 'unrolled' cylindrical surface these
lines of constant phase are straight.
\subsubsection{Taylor vortex like patterns --- $m$=0}
 For rotationally symmetric Taylor vortex like perturbations the line pattern of
 constant phases, $kz_0 = n 2 \pi$, is parallel to ${\bf e}_\varphi$. The $m$=0
 pattern is stationary for $Re=0$. Only for finite through-flow it
 propagates axially with phase velocity
 \begin{equation} \label{w_phase_TVF}
w_{phase} = \frac{\omega}{k}
\end{equation}
that is proportional to $Re$. The main reason for this is that the azimuthal 
flow of the basic
CCF state is precisely parallel to the vortex lines of constant phase
while the APF flow being perpendicular to them can advect them. The latter
happens in our axially periodic system that does not exert any phase pinning 
at the axial boundaries as soon as $Re > 0$. 
\subsubsection{Spiral patterns --- $m \neq 0$}
The vortex modes (\ref{Ansatz}) with axial wave number $m \neq 0$ have spiral
structure. When $m/k$ is positive (negative) the lines of constant phase
$z_0(\varphi,t)$ (\ref{z_phase}) wind in a left spiral L-SPI (right spiral 
R-SPI) around the cylindrical surface $r=const$ with negative (positive) slope
$\partial_{\varphi} z_0 = -m/k$. The lines of constant phase and with it the
whole spiral stucture rotates in $\varphi$ rigidly with angular velocity
  \begin{equation} \label{phidot_phase}
\dot{\varphi}_{SPI} = \frac{\omega}{m} \, .
\end{equation}
In the absence of an externally imposed through-flow, $Re$=0, this rotation
proceeds for L-SPI and R-SPI alike into the same direction as the rotation of 
the inner cylinder. The reason is that
the spiral perturbations are advected by the inner part of the azimuthal CCF
which is relevant for the centrifugal instability leading to vortex
generation. A model explaining this effect is presented in \cite{HOF-LUE}. 

There
are two immediate consequences of this advective origin of the spiral dynamics
induced by the inner cylinder's rotation: ({\it i}) With the latter being
by definition positive -- in this work the inner cylinder is taken to rotate in  
positive $\varphi$-direction -- also $\omega (k,m) /m $ is positive for $Re$=0.
Hence, say, an $m=1$ ($m=-1$) spiral has positive (negative) frequency for 
$Re$=0. ({\it ii}) Then a L-SPI (R-SPI) being defined by  $m/k > 0$ ($m/k < 0$)
propagates for $Re$=0 upwards (downwards) with positive (negative) axial phase 
velocity
\begin{equation} 
w_{phase} = \frac{\omega}{k} = \frac{m}{k} \dot{\varphi}_{SPI}
\end{equation}
that is directly related to its {\em positive} angular
velocity $\dot{\varphi}_{SPI}$.

An externally applied axial through-flow changes the axial phase velocities and
frequencies of the $m=0$ and $m \neq 0$ vortex modes roughly proportional to 
$Re$, i.e., 
\begin{equation} 
w_{phase}(k,m,Re) - w_{phase}(k,m,Re=0) \propto Re \, .
\end{equation}
Simultaneously the rotation rates, $\dot{\varphi}_{SPI} = w_{phase} k /m$, of
the spirals are changed accordingly. Thus, in each case the vortex frequencies 
are largely determined by the basic state's advective properties, i.e., by the 
combination of azimuthal advection by $v_{CCF}$ and axial advection by 
$w_{APF}$.
 \subsection{Symmetries} \label{SEC.SYMM-EIGENVALUE}
Here we consider symmetry properties of axially extended vortex perturbations of
the form (\ref{Ansatz}) with real wave number $k$. Symmetry relations between 
different vortex fronts with complex wave number $Q$ are discussed later on in 
Sec.~\ref{SEC.FRONT-SYMM}.

Table \ref{TAB.MOD-SYMM} shows the symmetry transformations that leave the
eigenvalue problem unchanged. They reflect ({\it i}) that ${\cal L}$ transforms
under complex conjugation (indicated by an overline) as 
\begin{equation} \label{Lcomplexconj}
\overline {{\cal L}(k,m,\sigma)} = {\cal L} (-k,-m, \overline {\sigma}) 
\end{equation}
and ({\it ii}) that the NSE (\ref{NAVIER2}) are invariant under an axial 
reflection $(z,Re,w) \longrightarrow (-z,-Re,-w)$.

Thus one infers, for example, that the growth rate (frequency) of the 
characteristic exponent $ \sigma = \gamma - i \omega$ for $m = 0$ vortices is 
an even (odd) function of $k$ and $Re$. For perturbations with $m\ne 0$ one 
finds that
\begin{eqnarray}
\gamma(k,m,Re) =  \gamma(-k,-m,Re) &=& \gamma(-k,m,-Re) =  \gamma(k,-m,-Re)  \\
\omega(k,m,Re) = -\omega(-k,-m,Re) &=& \omega(-k,m,-Re) = -\omega(k,-m,-Re)
\end{eqnarray}  
and that the spatiotemporal structure including the amplitude functions of a 
L-spiral perturbation ($k/m > 0$) at $Re > 0$ is the same as that of a R-spiral
($k/m < 0$) at $Re < 0$. Note, however, that any finite through-flow breaks 
the axial mirror symmetry between L- and R-spirals at $Re=0$ so that, among
others,
\begin{equation}
\gamma_R (- Re) = \gamma_L (Re) \ne \gamma_R (Re) = \gamma_L (- Re)
\end{equation} 
when $Re \ne 0$. But the symmetry relations are such that it
suffices to investigate, say, positive $k$ combined together with 
either ({\em i}) $m>0$ only for positive {\em and} negative $Re$ or, equivalently, 
({\em ii}) $m$ positive {\em and} negative for $Re>0$ only in order to get the
complete linear information on both spiral vortex types. 
\subsection{Bifurcation thresholds} \label{SEC.RESULTS1}
\begin{figure} 
 \includegraphics[clip=true,width=10cm,angle=0]{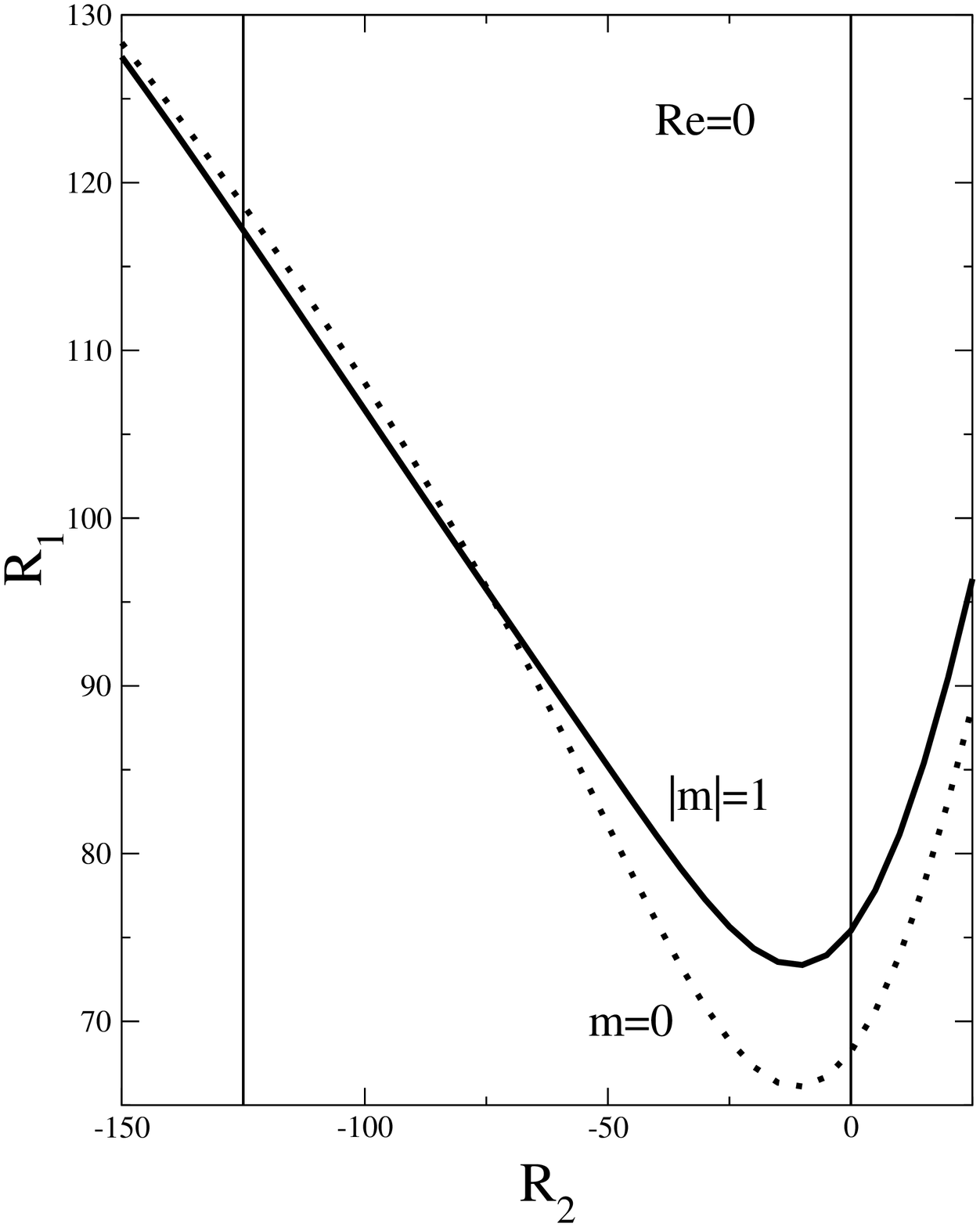}
 \caption{ Critical bifurcation thresholds $R_{1c}(R_2)$ for $m=0$ and 
 $m=\pm 1$ vortex patterns with the respective
critical wave numbers, $k_c(m, R_2)$, as functions of $R_2$ in the absence of
axial through-flow. The vertical lines mark the two 
representative outer Reynolds numbers $R_2 = 0$ and $R_2 = -125$ that are 
investigated in more detail in this work. The radius ratio is $\eta=0.5$.
\label{FIG.R1c_R2}}
 \end{figure}

Figure~\ref{FIG.R1c_R2} shows the {\em critical} bifurcation thresholds 
$R_{1c}(R_2)$ for $m=0$ and $m=\pm 1$ vortex patterns with the respective
{\em critical} wave numbers, $k_c(R_2)$, as functions of $R_2$ in the absence of
through-flow. 

\begin{figure}
 \includegraphics[clip=true,width=10cm, angle=0]{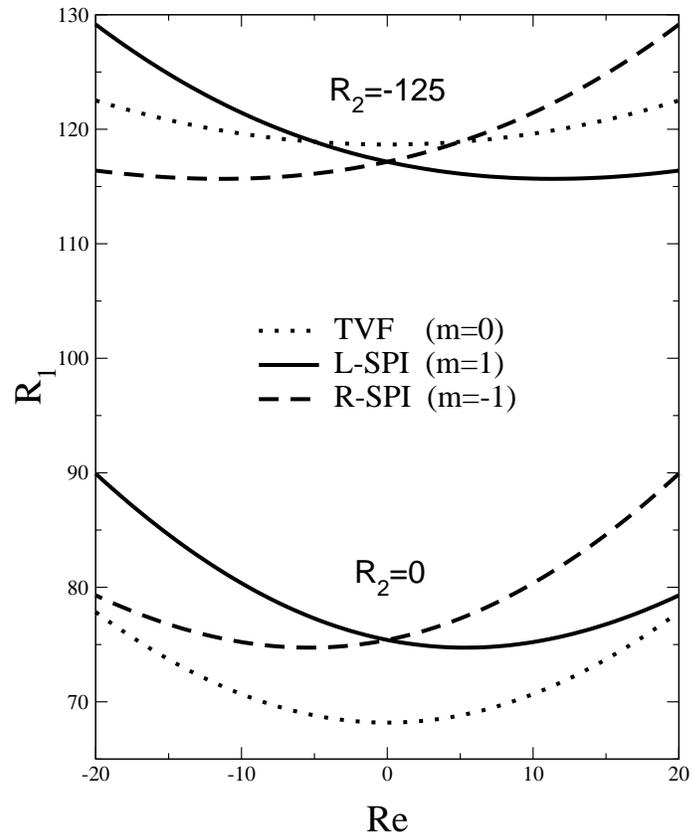}
 \caption{ Evolution of the critical bifurcation thresholds $R_{1c}(Re)$  
 for $m=0$ and $m=\pm 1$ vortex patterns 
with through-flow Reynolds number $Re$. The two outer Reynolds 
numbers $R_2 = 0$ and $R_2 = -125$ are marked in Fig.~\ref{FIG.R1c_R2}. 
The radius ratio is $\eta=0.5$. 
\label{FIG.R1c_Re}}
 \end{figure}

The vertical lines in Fig.~\ref{FIG.R1c_R2} mark the two outer
Reynolds numbers $R_2 = 0$ and $R_2 = -125$ for which we show in 
Fig.~\ref{FIG.R1c_Re} as representative examples how the critical thresholds
evolve with through-flow Reynolds number $Re$. The above discussed
symmetry relation $\gamma_R (- Re) = \gamma_L (Re)$ between the growth rates 
of R- and L-spirals implies the corresponding relation between the respective 
bifurcation thresholds (full and dashed lines in Fig.~\ref{FIG.R1c_Re}).
In the remainder of this paper we therefore restrict ourselves without loss of
information to positive $Re$.

For small $Re>0$ the axial flow stabilizes the basic state against growth of
TVF ($m=0$) and R-SPI ($m=-1$) perturbations. On the other hand, the 
bifurcation threshold for 
L-SPI ($m=1$) vortex patterns that propagate into the same direction as the
through-flow decreases at small $Re$ and increases only at larger $Re$. 
The upwards shift of the $m=0$ threshold with increasing $Re$ is 
stronger than the one for $m=1$. Thus, eventually the latter comes to lie below
the former and consequently the growth of L-spirals propagating into the same
direction as the through-flow is favored for sufficiently large $Re$ even when
$R_2=0$.

\begin{figure} 
\includegraphics[clip=true,width=9cm, angle=0]{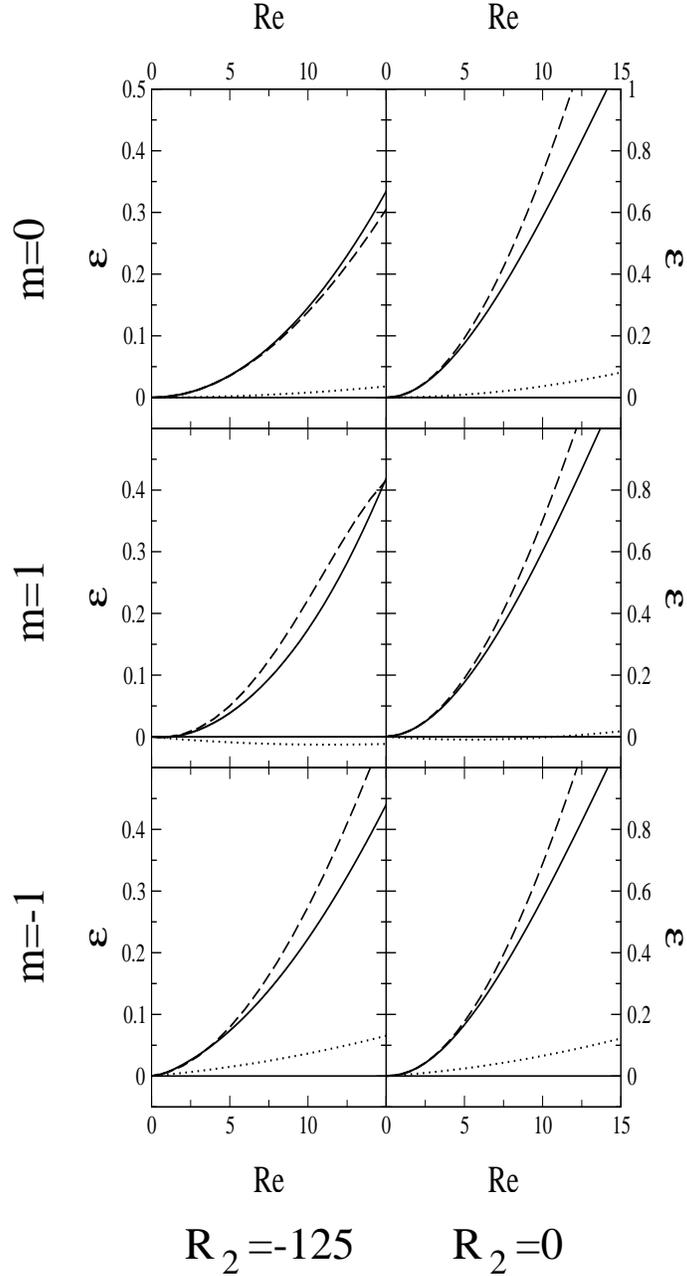}
 \caption{Stability boundaries of the basic flow state as functions of the
 through-flow Reynolds number $Re$. Dotted lines show the reduced critical 
 bifurcation thresholds 
 $\epsilon_c$ (\ref{epsilon_c}) for axially extended vortex patterns.  
 The full (dashed) boundary lines, $\epsilon_{c-a}$, between the 
 convectively and absolutely unstable parameter regions for vortex growth 
 were obtained from the eigenvalues of the full NSE (the GLE approximation) --
 cf. Sec.~\ref{SEC.LOCAL-PERTUB}. The radius ratio is $\eta=0.5$.
 \label{FIG.EPSILON}}
 \end{figure}

Dotted lines in Fig.~\ref{FIG.EPSILON} show the reduced critical threshold
curves $\epsilon_c(Re)$ (\ref{epsilon_c}) as functions of $Re$ for the two
representative outer Reynolds numbers $R_2=0$ and $R_2=-125$. The other lines
in Fig.~\ref{FIG.EPSILON} are discussed in Sec. \ref{SEC.LOCAL-PERTUB}. Our
numerical results for $\epsilon_c(Re)$ that were obtained in steps of 
$\delta Re = 1$ were fitted in the range $Re=-20\cdots20$ to the following 
expression
\begin{equation} \label{fitfunction}
f=a_{1}Re+a_{2}Re^{2}+a_{3}Re^{3}+a_{4}Re^{4} \, .
\end{equation}
The fit parameters $a_n$ are listed for different $R_2$ in Tables 
\ref{TAB.FITPARA} and \ref{TAB.FITPARA2} for TVF ($m=0$) and L-SPI $(m=1)$,
respectively. The threshold curves for R-SPI $(m=-1)$ are obtained according to
Sec.~\ref{SEC.SYMM-EIGENVALUE} from those for L-SPI by $Re \rightarrow -Re$, 
i.e., by changing the sign of the odd coefficients in Table~\ref{TAB.FITPARA2}.
 \section{Localized vortex perturbations} \label{SEC.LOCAL-PERTUB}
So far we have considered axially extended vortex perturbations of the form
$\psi(r,\varphi,z,t) = \phi(r)e^{i(kz + m\varphi)}$
which are single eigenmodes of the operator ${\cal L}$~(\ref{Result.Operator}).
For supercritical control parameters, $R_1 > R_{1c}(m)$, a finite band of axial
wave numbers can grow and with it also axially localized wave packets 
consisting of superpositions of vortex modes.
\subsection{Vortex packets} \label{SEC.PACKETS}
Let us consider first an infinitesimal initial perturbation with azimuthal 
wave number $m$ that is axially localized, i.e., a wave packet that
consists of a superposition of vortex modes of different $k$ but common $m$ --
an initial perturbation containing different $m$-modes would be just a sum of
the above described vortex packets that would evolve independently of each other
as long as the linear description is valid.
 
After fast transients have decayed a pulse like perturbation survives
with axial wave numbers within the unstable band centered around the wave 
number of maximal growth $k_{max}(m) \simeq k_c(m)$. Since the above described
wave packet contains vortex modes that can grow the pulse will grow as well. 
Simultaneously the pulse center travels axially for small $\epsilon$ with 
the critical goup velocity
\begin{equation}
v_{g}=\left. \frac{\partial \omega(k)}{\partial k} \right|_{k_{c}}.
\end{equation}
Hence when at a bifurcation threshold the frequency is nonzero with a finite 
group velocity then the supercritical spatiotemporal growth behavior of an 
axially {\it localized} vortex perturbation differs significantly from an axially 
{\it extended} vortex mode. The growth of the latter is axially uniform which is
not the case for the former. 

Furthermore, 
one has to distinguish between two different supercritical regimes for the 
former: ({\em i}) In the so-called convectively unstable parameter regime 
of the basic state the vortex packet moves with the velocity $v_{g}$ faster 
away than it grows -- while growing in the frame comoving with $v_{g}$ the 
pulse moves out of the system so that the basic state is restored 
\cite{b&b,HUE}. In other words, the two fronts that join 
the pulse intensity envelope to the structureless state propagate 
both in the direction in which the packet center moves.
{\em(ii)} In the so called absolutely unstable parameter regime 
the growth rate of the packet is so large that one front propagates in the 
laboratory frame opposite to the center motion. Thus, the packet expands not 
only into the direction of the pulse motion but also opposite to it 
\cite{b&b} so that eventually the initial perturbation can fill the entire 
system. But the linear growth analysis of the vortex fields does not 
determine in what nonlinear final stable state the system will end nor what 
possible intermediate nonlinear transient behavior might occur.

However, this analysis has an important implication for experiments with 
through-flow: 
Developed vortex structures can be seen in finite systems with vortex 
suppressing inlet conditions only in the absolutely unstable regime which is 
typically realized at larger $R_1 > R_{1c}$ (cf. Fig.~\ref{FIG.EPSILON}) if one 
leaves aside noise-sustained patterns  \cite{Deissler85,BAC-Pre94} in the 
convectively unstable regime. In this latter regime the vortex front that 
connects to the zero amplitude inlet condition moves downstream (we assume that
the fronts of our forwards bifurcating vortex structures are linearly selected 
thus excluding the buildup of nonlinear fronts that might revert their 
propagation direction). In the absolutely unstable regime, on the other hand, 
an upstream  motion of this front is stopped by the  inlet condition at a
characteristic downstream growth length from the inlet. This growth length
of the downstream evolving vortex structure diverges \cite{MLK89,BLRS96} when 
approaching the boundary between convective and absolute instability from the 
latter regime. 
 
It is remarkable that based on the experimental observations of Takeuchi and 
Jankowski \cite{TAK-JAN} such a behavior was discussed already in 1979 [cf. 
figures 4 and 5(a) and their discussion in Sec. 6 of ref.~\cite{TAK-JAN}], albeit
without invoking the concept of absolute and convective instability \cite{b&b} 
which was introduced to a broader fluid dynamics community only a few years 
later \cite{HUE}.

 \subsection{Boundary between convective and absolute instability: 
 saddle point analysis} \label{SEC.SADDLE-POINT}
The boundary between convective and absolute instability of the basic flow 
 against growth of vortices with a particular azimuthal wave number $m$ 
 is marked by those parameter combinations for which one of the fronts of the 
 linear packet of $m$-vortices reverts its propagation direction in the 
 laboratory  frame: In the convectively unstable parameter regime this front 
 propagates in the same direction as the
 center of the packet, in the absolutely unstable regime it moves opposite to
 it, and right on the boundary between the two regimes the front is stationary
 in the laboratory frame. 
 
 This parameter combination can be determined 
by a saddle point analysis of the linear complex dispersion relation
$\sigma(Q)$ over the plane of complex wave numbers \cite{HUE,CRO-HOH}
\begin{equation}
Q = \Re Q + i\Im Q = k - i K \, . 
\end{equation}
Here we do not display the dependence of 
\begin{equation}
\sigma(Q) = \Re \sigma(Q) + i\Im \sigma(Q) = \gamma(Q) - i \omega(Q) 
\end{equation}
on the parameters $R_1, R_2, Re, \eta$ and we also do not indicate that the
dispersion relations for vortex perturbations with different azimuthal wave
numbers $m$ are different. The boundary condition of no growth for a front that 
is stationary in the laboratory system is  
\begin{equation} \label{Wachsbed.}
\Re\sigma = 0
\end{equation}
at the appropriate saddle position, $Q^\ast$, of $\sigma(Q)$ in the complex
wave number plane \cite{HUE}. Here $Q^\ast$ follows from 
\begin{equation} \label{saddle}
\frac{d\sigma(Q)}{dQ}\Big|_{Q^\ast} = 0.  
\end{equation}
For fixed $m, R_2, Re,\eta$ eqs.(\ref{Wachsbed.},\ref{saddle}) yield 
$R_{1c-a}$. Here and in the following the subscript $c-a$ identifies
boundaries between convective and absolute instability. Thus, e.g., the basic 
flow state is convectively [absolutely] unstable against vortex perturbations 
with azimuthal wave number $m$ for $R_{1c}(m) < R_1 < R_{1c-a}(m)$ 
[$R_1 > R_{1c-a}(m)$] and absolutely stable when $R_1 < R_{1c}(m)$. 

We are interested here in the $Re$ dependence of these thresholds and we will 
discuss to that end the reduced boundary quantities 
\begin{equation}
\mu_{c-a}(Re)=\frac{R_{1c-a}(Re)}{R_{1c}(Re)}-1
\end{equation}
and
\begin{equation}
\epsilon_{x}(Re)=\frac{R_{1x}(Re)}{R_{1c}(Re=0)}-1,
\ \textrm{where}\ x=c\ \textrm{or}\ x=c-a \, .
\end{equation}
Because of the symmetries of $\sigma(Q)$, of its saddle-point $Q^\ast$, and of
the resulting boundary $R_{1c-a}(m, Re)$ one has
\begin{eqnarray}
k^{\ast}(-m,-Re)&=&k^{\ast}(m,Re)\\
K^{\ast}(-m,-Re)&=&-K^{\ast}(m,Re)\\
\mu_{c-a}(-m,-Re)&=&\mu_{c-a}(m,Re)\\
\epsilon_{c-a}(-m,-Re)&=&\epsilon_{c-a}(m,Re)
\end{eqnarray}
so that it suffices again to investigate only positive through-flow.

 \subsection{Numerical procedures} \label{SEC.NUM-PROC2}
In order to determine the boundary between convective and 
absolute instability via the solution of eqs.~(\ref{Wachsbed.}, \ref{saddle})
one has to evaluate the dispersion relation $\sigma(Q)$ for complex $Q$ 
\cite{TAG-EDW-SWI,BAC-Pre94}. 
To that end we solved the eigenvalue problem (\ref{resultEW}) of
the full field equations for {\em complex} $Q$ (cf., Sec.~\ref{SEC.NSE1}). 
In addition
and for comparison we used for $\sigma(Q)$ the Ginzburg-Landau amplitude 
equation approximation which only requires knowledge of 
$\sigma(k)$ along the real $k$-axis.  
 \subsubsection{Ginzburg-Landau amplitude equation approximation} 
 \label{SEC.GLA1} 
Within this approximation the dispersion relation $\sigma(Q)$ for vortex 
modes is expanded in $Q$ and $R_1$ around the critical point $k_c, R_{1c}$
\begin{eqnarray}
\sigma(Q,R_1)&=&\sigma_c+\left(\frac{\partial \sigma}{\partial Q}
\right)_c(Q-k_c)+\frac{1}{2}\left(\frac{\partial^2 \sigma}{\partial
Q^2} \right)_c (Q-k_c)^2\nonumber\\
& &+\left(\frac{\partial \sigma}{\partial R_1}
\right)_c(R_1-R_{1c}) + h.o.t. \\
&=&-i\omega_c-iv_g(Q-k_c)  
-\frac{\xi_0^2}{\tau_0}(1+ic_1)(Q-k_c)^2 \nonumber\\
& &+ \frac{(1+ic_0)}{\tau_0}\mu + h.o.t. \, . \label{Ginz-Landau-Naeherung}
\end{eqnarray}
The expansion coefficients in the above expressions appear also in the linear 
parts of the Ginzburg-Landau amplitude equation \cite{CRO-HOH}. They are 
obtained from the numerical solution of the eigenvalue problem (\ref{resultEW}) 
of the full field equations for real $k$ close to the 
critical point \cite{REC-LUE-MUE,PIN}.  

Within the GLE approximation one gets from eqs.~(\ref{Wachsbed.}, \ref{saddle})
\begin{equation} \label{QstarGLE}
 Q^\ast=\underbrace{k_c-\frac{c_1 \tau_0 v_g}{2(1+c_1^2)\xi_0^2}}_{k^\ast}
 -i \underbrace{\frac{\tau_0 v_g}{2(1+c_1^2)\xi_0^2}}_{K^\ast},
\end{equation}
\begin{equation}
\mu_{c-a}=\frac{v_{g}^{2}\tau_{0}^{2}}{4(1+c_{1}^{2})\xi_{0}^{2}}, 
\end{equation}
and
\begin{equation}
\epsilon_{c-a}(Re)= \epsilon_{c}(Re) + 
\frac{v_g^2\tau_0^2}{4(1+c_1^2)\xi_0^2} \left[1 + \epsilon_c(Re)\right].
\end{equation}
Note that the GLE coefficients $v_g, \tau_0, \xi_0, c_0, c_1$ depend on 
$m, R_2, Re, \eta$. For $m=0$ and $R_2=0$ also the nonlinear coefficients of the
GLE have been obtained for several $\eta$ as functions of $Re$ \cite{REC-LUE-MUE}.

 \subsubsection{Dispersion relation $\sigma(Q)$ of the NSE for complex $Q$} 
 \label{SEC.NSE1}
To assess the quality of the GLE results for the boundary between convective and
absolute instability we determined the dispersion relation of the NSE not only
for real $k$ close to the critical point but for complex axial wave numbers $Q$
that lie in the vicinity of the relevant saddle locations $Q^\ast$ of 
$\sigma(Q)$. Having determined $\sigma(Q)$ as described in 
Sec.~\ref{SEC.NUM-PROC1} we solved the equations (\ref{Wachsbed.},\ref{saddle})
for $R_{1c-a}$ in the form
\begin{equation}
\gamma (Q^\ast)=0 ; \quad
\frac{\partial \gamma(Q)}{\partial K} \Big|_{Q^\ast}=0 ; \quad
\frac{\partial \gamma(Q)}{\partial k} \Big|_{Q^\ast}=0
\end{equation}
that follows from using the Cauchy-Riemann relations for $\sigma(Q)$.

 \subsection{Results} \label{SEC.RESULTS2}

\begin{figure}
 \includegraphics[clip=true,width=10cm, angle=0]{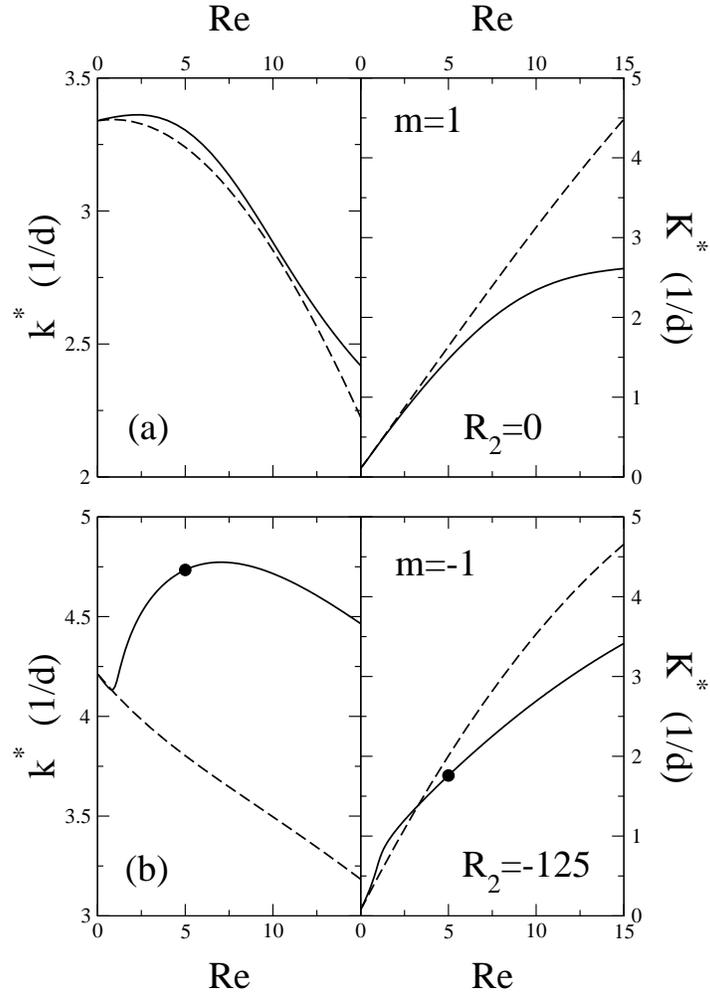}
 \caption{Evolution of the coordinates $k^\ast$ and $K^\ast$ of the saddle point
  that determines the boundary, $R_{1c-a}$, between convective and absolute 
  instability with through-flow Reynolds number $Re$. Full (dotted) lines are
  evaluated with the correct NSE (approximate GLE) dispersion relation. Filled
  circles in (b) at $Re=5$ 
  mark the coordinates of the saddle in 
  Fig.~\ref{FIG.2EW}. The radius ratio is $\eta=0.5$. 
 \label{FIG.SATTEL}}  
 \end{figure}

In Fig.~\ref{FIG.SATTEL} we show the $Re$-variation of the real and imaginary
parts of the relevant saddle-point $Q^\ast = k^\ast - i K^\ast$ at the boundary  
$R_{1c-a}$ for two characteristic cases: (a) L-SPI ($m=1$) at $R_2=0$ and (b) 
R-SPI ($m=-1$) at $R_2=-125$. In each case full (dashed) lines were obtained
from the correct NSE (approximate GLE) dispersion relation $\sigma(Q)$. 

Case (a) is representative for a situation where the GLE approximation
reasonably well reproduces the correct result -- at least for small $Re$ -- and
starts to deviate significantly only for larger $Re$. On the other hand, in case
(b) the GLE approximation to $k^\ast$ displays a smooth variation with $Re$
that reflects the smooth variation of the saddle $Q^\ast_{GLE}$ (\ref{QstarGLE})
of $\sigma_{GLE}(Q)$ (\ref{Ginz-Landau-Naeherung})
while the real part $k^\ast$ of the saddle location of the correct dispersion
relation undergoes a dramatic change around $Re \simeq 0.8$. The reason is that
the surfaces of $\gamma(Q) = \Re \sigma(Q)$ over the $Q$-plane for the two 
largest eigenvalues intersect and change their order along the $\gamma$ axis. 
Thus, the saddle that is relevant for the $c-a$ transition and that has the 
largest $\gamma$ value switches from one eigenvalue surface to another. Similarly
the eigenvalue surface might have another saddle and their ordering
in $\gamma$ changes.
In contrast to that the GLE approximation produces only a
single eigenvalue tracing out the smooth surface $\sigma_{GLE}(Q)$ 
(\ref{Ginz-Landau-Naeherung}).

\begin{figure}
 \includegraphics[clip=true,width=10cm, angle=0]{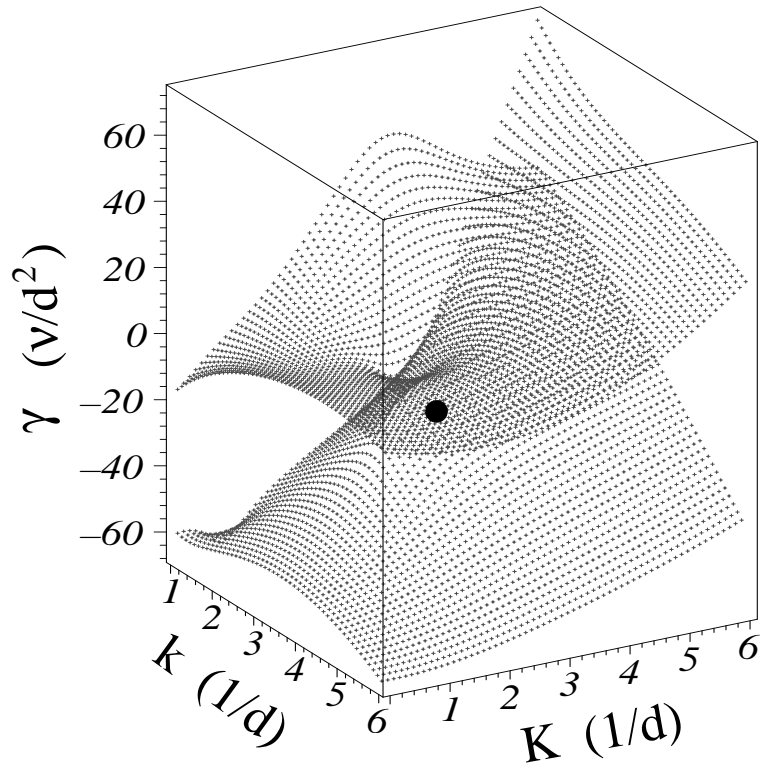}
 \caption{Real parts $\gamma (Q)$ of the two biggest eigenvalues of the NSE over
 the complex $Q-$plane for $m=-1$, $R_2=-125$, $Re=5$, 
 $\mu=\mu_{c-a}$, $\eta=0.5$. The filled circle  
 marks the saddle at 
 $\gamma (Q^\ast) = 0$ that determines the boundary beteen convective and
 absolute instability. 
 \label{FIG.2EW}}  
 \end{figure}

\begin{figure}
 \includegraphics[clip=true,width=10cm, angle=0]{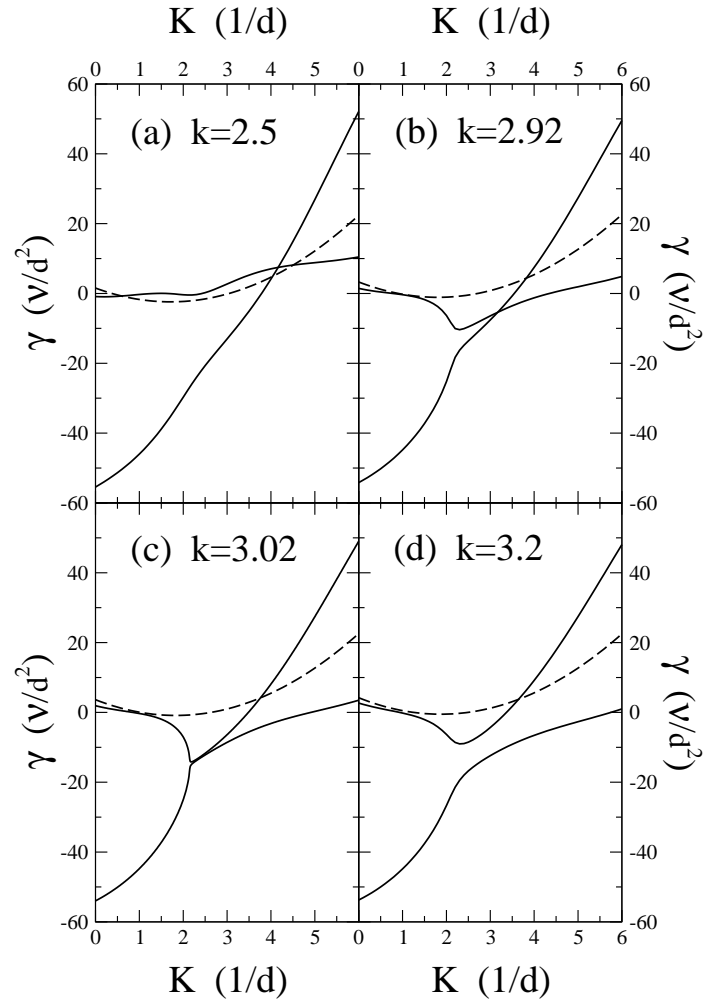}
 \caption{Full lines are sections through the surfaces of $\gamma (Q)$ 
 (Fig.~\ref{FIG.2EW}) for the two biggest eigenvalues of the NSE at constant 
 values of $k$ in the vicinity of the intersection of the surfaces. Dashed
 lines show $\gamma (K)$ obtained from the GLE approximation to the dispersion
 relation. Parameters are $m=-1$, $R_2=-125$, $Re=5$, $\mu=\mu_{c-a}$, 
 $\eta=0.5$. 
 \label{FIG.SATTEL-SCHNITTE}}  
 \end{figure}

To give an impression of such an intersection of the correct dispersion 
surfaces we show in Fig.~\ref{FIG.2EW} their real parts over the complex 
$Q-$plane at $Re=5,\mu=\mu_{c-a}$. Full lines in Fig.~\ref{FIG.SATTEL-SCHNITTE} 
show different constant-$k$ sections through them in the intersection range. 
At $Re=5,\mu=\mu_{c-a}$ the saddle has moved already away from the
intersection region. The saddle coordinates 
($Q^\ast \simeq 4.73 -i 1.76, \gamma^\ast=0$) are indicated in
Figs.~\ref{FIG.SATTEL}, \ref{FIG.2EW} by full dots.

\begin{figure}
 \includegraphics[clip=true,width=10cm, angle=0]{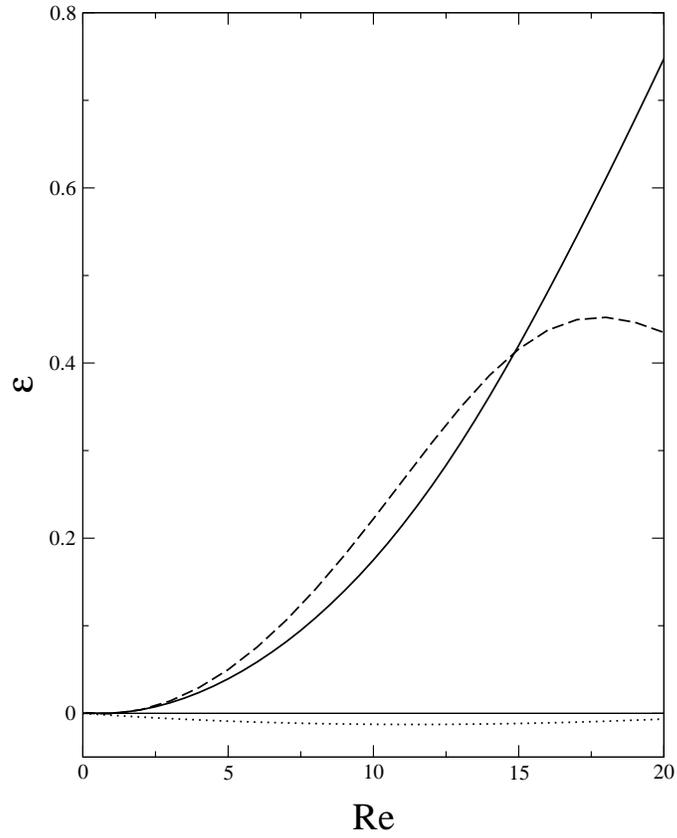}
 \caption{Stability boundaries of the basic flow state as functions of the
 through-flow Reynolds number $Re$ for $m=1$, $R_2=-125$. The full (dashed) 
 boundary lines, $\epsilon_{c-a}$, between the 
 convectively and absolutely unstable parameter regions for vortex growth 
 were obtained from the eigenvalues of the full NSE (the GLE approximation) --
 cf. Sec.~\ref{SEC.LOCAL-PERTUB}. Dotted lines show the reduced critical 
 bifurcation thresholds 
 $\epsilon_c$ (\ref{epsilon_c}) for axially extended vortex patterns.  
 The radius ratio is $\eta=0.5$.
 \label{FIG.EPSILON-RE=20}}  
 \end{figure}

While the saddle locations $Q^\ast_{GLE}$ of the GLE approximation 
$\sigma_{GLE}(Q)$ (\ref{Ginz-Landau-Naeherung}) can differ substantially from
the saddle $Q^\ast$ of the correct dispersion relation the difference in the
boundaries between convective and absolute instability is typically much less
pronounced -- cf, e.g., the dashed and full curves for $\epsilon_{c-a}$ in  
Fig.~\ref{FIG.EPSILON}. There the GLE results (dashed lines) agree in each case
quite well with the correct boundaries (full lines) up to, say $Re =5$.
However, as a consequence of the typical increase of the reduced boundary
$\mu_{c-a}$ with increasing $Re$ the quality of the GLE results for 
$\mu_{c-a}$ generally
deteriorates. After all the GLE is strictly valid only for $\mu \to 0$.
Fig.~\ref{FIG.EPSILON-RE=20} shows an example ($m=1, R_2=-125$) where the GLE
prediction for the boundary shows even a qualitative different variation with 
$Re$ for, say, $Re>15$. We have observed similar behavior -- partly at larger
$Re$ -- also for other combinations of $m, R_2$. 

Tables~\ref{TAB.FITPARA} and \ref{TAB.FITPARA2} contain the coefficients for
fitting our results for $\epsilon_{c-a}(m=0, R_2, Re)$ and 
$\epsilon_{c-a}(m=1, R_2, Re)$ in the same way as described in 
Sec.~\ref{SEC.RESULTS1}. Also here the  
boundary curves for R-SPI $(m=-1)$ are obtained according to
Sec.~\ref{SEC.SYMM-EIGENVALUE} from those for L-SPI $(m=1)$ by 
$Re \rightarrow -Re$, i.e., by changing the sign of the odd coefficients in 
Table~\ref{TAB.FITPARA2}.
 \section{Fronts and pulses} \label{FRONTS}
The vortex fronts that we are investigating here and that appear also as 
constituents of vortex pulses are 
perturbations of the basic state where the fields (locally) have the form
\begin{equation} \label{Einhullende}
\psi  \sim e^{i (k^\ast z - \omega^\ast t)}
e^{K^\ast (z - v^\ast t)}e^{im\varphi}
\end{equation}
in the laboratory frame. This form describes long-time spatiotemporal
properties of uniquely selected linear fronts \cite{CRO-HOH}.
Axial front velocity $v^\ast$, axial growth rate $K^\ast$ of the front envelope,
wave number $k^\ast$ of the vortex pattern under the front, and its frequency 
$\omega^\ast$ are determined 
by the saddle behavior of the linear complex dispersion relation 
$\sigma(Q,m) = \gamma (Q,m) - i \omega (Q,m)$ of the field equations over the 
plane of complex axial wave numbers $Q = k - i K$. Since superpositions of vortex
perturbations with different $m$ evolve independently from each other within the
linear description as mentioned already in Sec.~\ref{SEC.PACKETS} we consider 
here only fronts of vortex perturbations that have a common azimuthal wave 
number $m$ and we do not always display the latter explicitly. 

The saddle condition is \cite{CRO-HOH} 
\begin{equation} \label{saddlecondition}
\frac{d}{dQ} \left[\sigma \left(Q\right)  \: \: + \: \: i \: v \: Q \right] 
\Big|_{Q=Q^\ast}  =  0 .
\end{equation} 
And the stationarity requirement that the temporal growth rate of the front 
vanishes in the frame comoving with the front velocity $v^\ast$ demands that 
\begin{equation}
\label{gammasaddle}
0 = \Re \left[ \sigma (Q) + i v Q \right] \Big|_{Q = Q^\ast}
= \gamma (Q^\ast) + v^\ast K^\ast .
\end{equation}
We combine (\ref{saddlecondition}) and (\ref{gammasaddle}) into the three 
equations 
\begin{equation} \label{Bedingungen2}
v^\ast = - \frac{\gamma (Q^\ast)}{K^\ast}
= - \frac{\partial \gamma (Q)}{\partial K} \Big|_{Q^\ast} \quad ; \quad
\frac{\partial \gamma (Q)}{\partial k} \Big|_{Q^\ast} = 0 
\end{equation}
that we have solved for $v^\ast, k^\ast, K^\ast$.

 \subsection{Notation} \label{SEC.FRONT-NOTATION}

\begin{figure}
 \includegraphics[clip=true,width=10cm, angle=0]{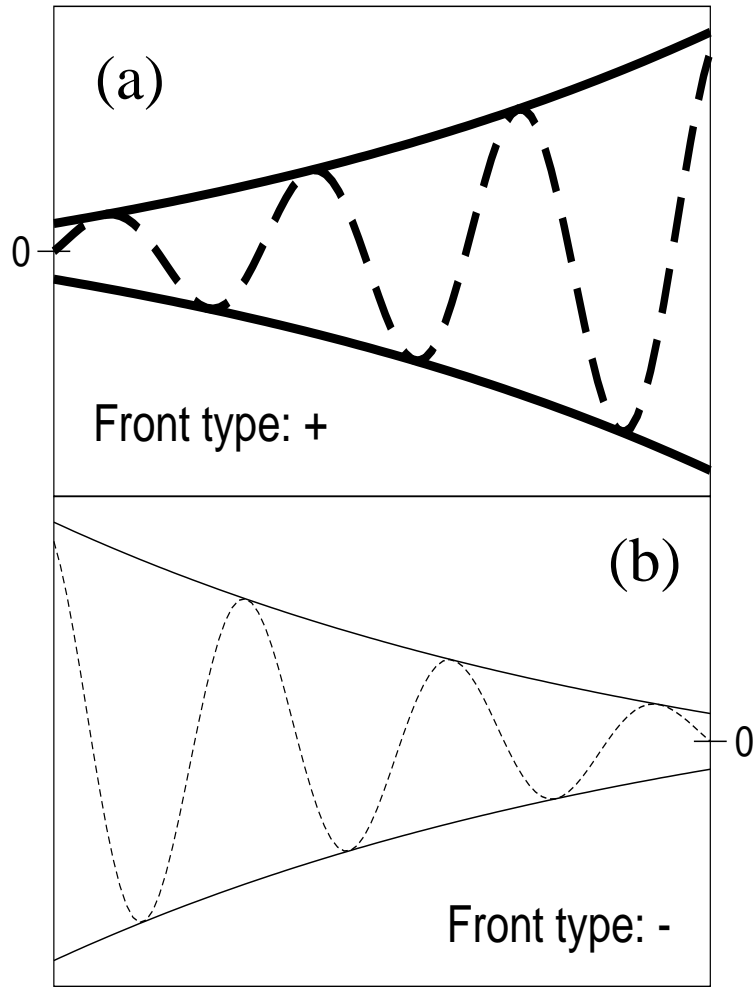}
 \caption{Schematic plot of different vortex fronts. Thick full lines in (a) and 
 thin full lines in (b) show intensity envelopes  of $+$ type and of $-$ type 
 fronts connecting to the basic state  at $z \to -\infty$ and at 
 $z \to \infty$, respectively. This line convention --- thick ones for $+$ type 
 fronts and thin ones for $-$ type fronts --- is used also in 
 Fig.~\ref{FIG.FRONTEN}. Dashed lines indicate the vortex field 
 growing under the front.
 \label{FIG.FRONTDIRECT}}  
 \end{figure}

The front envelope of (\ref{Einhullende}) varies axially with $e^{K^\ast z}$. 
If $K^\ast>0$, then the perturbation 
grows at $z=-\infty$ out of the basic state. We call such a front to be of type 
$+$ and identify the associated front
properties by a subscript $+$. On the other hand, for $K^\ast<0$ we have a front 
of type $-$ with an intensity envelope that joins at $z=\infty$ with the basic 
state. So the two subscripts $\pm$ identify
the axial variations of the front envelopes. A pulse-like perturbation of the 
basic state would consist suffiently away
from its center of a $+$ front for $z \to -\infty$ and of a $-$ front 
for $z \to \infty$. Correspondingly the eqs.~(\ref{Bedingungen2}) have two
different solutions: one of them describes a 
$-$ front and the other one a $+$ front. A schematic plot of the
different envelope types can be seen in Fig. \ref{FIG.FRONTDIRECT}.

We identify the vortex pattern 
that is unfolded under a front either by its azimuthal wave number $m$ or by
the superscripts T, L, R. Here T refers to a TVF-like pattern of vortices with
$m=0$, L denotes L-spiral vortices with $m=1$, and R identifies R-spiral vortices
with $m=-1$. Hence $m$ or equivalently the 
superscripts identify the spatiotemporal structure of the vortex pattern growing
under the front. In this work we restrict ourselves to these three vortex 
varieties. Since they appear under the above
described two front envelopes there are six different fronts that we have
investigated here.

 \subsection{Results} \label{SEC.RESULTS3}

In Fig.~\ref{FIG.FRONTEN} the results of our investigations are shown for 
fronts with azimuthal wave numbers $m$ = 0, 1, and -1 for axial
through-flow Reynolds numbers $Re$ =0, 10, and 20. In each case the outer 
cylinder is at rest, $R_2=0$. The front properties are presented as functions of
$\mu$. Within each 2$\times$2 block of figures (a)-(i) the left column shows 
real part $k^\ast$ and negative imaginary part $K^\ast$ of the saddle point and 
the right column shows the front velocity $v^\ast$ and the frequency 
$\omega^\ast$, respectively. They all start at $\mu = 0$ since vortex growth is 
possible only above the critical threshold, i.e., for $R_1 > R_{1c}(Re)$.
The two {\em critical} fronts at this threshold are degenerate ($k^\ast_- = 
k^\ast_+ = k_c, \, \omega^\ast_- = \omega^\ast_+ = \omega_c, \, 
v^\ast_- = v^\ast_+ = v_g$) with vanishing axial growth rates 
$K^\ast_- = K^\ast_+ = 0$. For $\mu>0$, however, $+$ and $-$ fronts differ from
each other thus reflecting differences in the respective saddle points.

The variation of $k^\ast, K^\ast, v^\ast$, and $\omega^\ast$ with $\mu$ and 
$Re$ is best understood by
comparison with the corresponding GLE approximation (cf. Sec.~\ref{SEC.GLA2}) and
by invoking symmetry relations (cf. Sec.~\ref{SEC.FRONT-SYMM}). We therefore
continue the discussion of our results in the following two sections 
\ref{SEC.GLA2} and \ref{SEC.FRONT-SYMM}. 

 \subsubsection{Ginzburg-Landau amplitude equation approximation} \label{SEC.GLA2}

\begin{figure}
 \includegraphics[clip=true, width=13cm, angle=0]{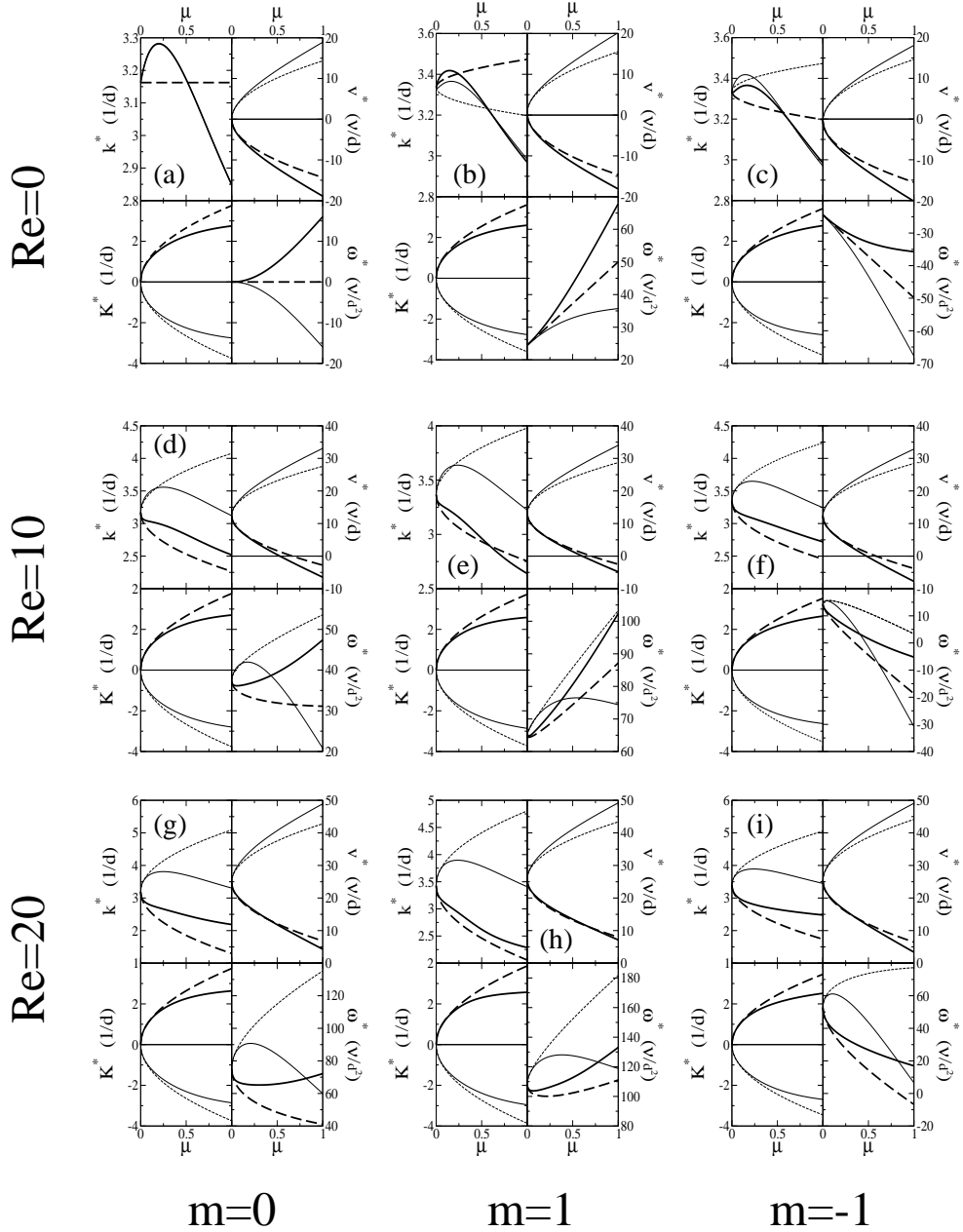}
 \caption{Front properties of vortices with azimuthal wave numbers $m=0, 1, -1$ 
 in systems with and without through-flow $Re$ for $R_2$ = 0, $\eta=0.5$ as 
 functions of  $\mu = R_1/R_{1c}(Re) -1$. Within each
 2$\times$2 block of figures (a)-(i) the left column shows the axial wave number
 $k^\ast$ and the axial growth rate $K^\ast$. The right column shows the front 
 velocity $v^\ast$ and the frequency $\omega^\ast$ in the laboratory frame. 
 Thick lines correspond to $+$ fronts, thin lines to $-$ fronts, respectively.
  Full (dashed) lines result from the saddle point analysis of the  
  dispersion relation of the NSE (GLE approximation). 
 \label{FIG.FRONTEN}}  
 \end{figure}

The saddle point analysis (\ref{saddlecondition} - \ref{Bedingungen2}) of the
GLE approximation (\ref{Ginz-Landau-Naeherung}) yield the following
front properties
\begin{eqnarray} 
K^{\ast}_{\pm}&=&\pm \sqrt{\frac{\mu}{(1+c_{1}^{2})\xi_{0}^{2}}},\label{K-GLE} \\
k^{\ast}_{\pm}&=&k_{c}-c_{1}K^{\ast}_{\pm},\label{k-GLE}\\
v^{\ast}_{\pm}&=&v_{g}-2(1+c_{1}^{2})\frac{\xi_{0}^{2}}{\tau_{0}}K^{\ast}_{\pm},
\label{v-GLE}\\
\gamma^{\ast}_{\pm}&=&-v^{\ast}_{\pm}K^{\ast}_{\pm},\label{gamma-GLE}\\
\omega^{\ast}_{\pm}&=& \omega_c +v_g (k^\ast_\pm - k_c)
+(c_1-c_0)\frac{\mu}{\tau_0} \label{omega-GLE}\, .
\end{eqnarray}
for the two fronts with $K_+^{\ast} >0 $ and $K_-^{\ast} < 0 $, respectively.
{\em All} quantities appearing in (\ref{K-GLE} - \ref{omega-GLE}) depend on 
whether they refer to T ($m$ = 0), L ($m$ = 1), or R ($m$ = -1) vortex fronts.
These GLE results are shown in Fig.~\ref{FIG.FRONTEN} by dashed lines.

They reasonably
well describe the small-$\mu$ behavior of the correct front properties 
(full lines in Fig.~\ref{FIG.FRONTEN}) which
were obtained from the correct dispersion relation of the NSE: As predicted 
by the small-$\mu$ GLE approximation (\ref{K-GLE} - \ref{omega-GLE}) one finds
that for small $\mu$ the axial growth rates 
$K^\ast$ vary $\propto \sqrt \mu$, that consequently also $k^\ast - k_c$
and $v^\ast - v_g$ vary $\propto \sqrt \mu$, and that
$\omega^\ast - \omega_c$ can have in addition also a contribution 
$\propto \mu$ when $c_1 - c_0 \neq 0$. The latter is the case for $m=\pm 1$
irrespective of $Re$ and for $m=0$ if $Re \neq 0$.

Note that for TVF fronts with $m=0, Re = 0$ the GLE predicts $\omega^\ast = 0$
whereas the correct dispersion in Fig.~\ref{FIG.FRONTEN}(a) seems to show for 
small $\mu$ a variation of 
$\omega^\ast \propto \mu^2$ that is beyond the range of applicability of the 
GLE. Thus, under the linear part of a moving 
TVF front there should be a non zero phase propagation in the laboratory 
frame with phase velocity $\omega^\ast / k^\ast$. The analogous behavior was
found also for convection rolls in the Rayleigh-B\'enard system \cite{BUE-LUE}.

At the boundary between convective and absolute
instability, $\mu =\mu_{c-a}$, the velocity 
$v_+^{\ast}$ and the temporal growth rate $\gamma^{\ast}_+$ of the $+$ front 
vanish in the laboratory frame while $v_-^{\ast}$ is positive there
(and given by 2 $v_g$ within the GLE approximation). Note that
according to Fig.~\ref{FIG.EPSILON} and 
tables~\ref{TAB.FITPARA} and \ref{TAB.FITPARA2} $\mu_{c-a}$ is zero for $m=0,
Re=0$ and very small for $m= \pm 1, Re=0$. Only for large enough
through-flow $\mu_{c-a}$ becomes sizeable. 
In the convectively unstable regime $0 < \mu < \mu_{c-a}$ both, the $+$ front 
as well as the $-$ front of a vortex pulse move into the same downstream 
direction as the through-flow, $0 < v_+^{\ast} < v_-^{\ast}$. In the absolutely
unstable regime $\mu > \mu_{c-a}$, however, the $+$ front moves upstream and 
the $-$front moves downstream, $v_+^{\ast} < 0 < v_-^{\ast}$. 

In ref.~\cite{NG-TUR} it was remarked that the phase velocity of spiral patterns 
in axial flow through a system of radius ratio $\eta=0.95$ \cite{SNY}
deviates from the critical one, $\omega_c/k_c$, of axially extended vortex
perturbations. While we have done only 
calculations for $\eta=0.5$ our results of Fig.~\ref{FIG.FRONTEN} and of the
small-$\mu$ GLE approximation (\ref{K-GLE} - \ref{omega-GLE}) shed some light on
the existence of such deviations. Since the experimental vortex structures grow
in downstream direction under intensity fronts one strictly speaking would have 
to compare with the phase velocity under such fronts that connect in the 
absolutely unstable regime to the fully developed downstream vortex pattern. 
Such a nonlinear analysis has been done for $m=0$ patterns \cite{BLRS96} but not
for spirals. However, already the phase velocities $\omega_+^\ast /k_+^\ast$
under our linear $+$ fronts that grow in downstream direction differ from the 
corresponding critical phase velocities $\omega_c /k_c$ of axially extended 
patterns --- cf. Fig.~\ref{FIG.FRONTEN} and eqs.~(\ref{K-GLE} - \ref{omega-GLE}).

 \subsubsection{Symmetries} \label{SEC.FRONT-SYMM}
The front properties shown in Fig.~\ref{FIG.FRONTEN} are largely influenced by
the symmetry properties of the system {\em without} through-flow although 
a finite $Re$ changes them.

Invariance of the field equations under $z \to -z$ for $Re=0$ implies that 
stationary perturbations with
$\omega_{c}=0$ (TVF) under a $+$ front are mirror images of those under a $-$ 
front. This implies for $Re=0$ the symmetry relations
\begin{equation} \label{symmet1}
(K,\omega,v)_{+}^{\ast T}=-(K,\omega,v)_{-}^{\ast T};\quad 
k_{+}^{\ast T}=k_{-}^{\ast T}.
\end{equation}
These symmetry properties of the two front types of T-perturbations can be seen
in Fig. \ref{FIG.FRONTEN}(a).

Now consider L and R perturbations. Here the invariance of the field equations 
under  $z \to -z$ for $Re=0$ implies
first of all that a spatially extended L-SPI with uniform amplitude is the 
mirror image of a spatially extended R-SPI.
Furthermore, a L-SPI under a $+$ front with positive $K$ is symmetry degenerate 
with a R-SPI under a $-$ front with
negative $K$. Similarly a R-SPI under a $+$ type front is the mirror image of a
 L-SPI under a $-$ front. This implies for $Re=0$ the symmetry relations   
\begin{equation} \label{symmet2}
(K,\omega,v)_{+}^{*L}=-(K,\omega,v)_{-}^{*R};\quad k_{+}^{*L}=k_{-}^{*R},
\end{equation}
\begin{equation} \label{symmet3}
(K,\omega,v)_{+}^{\ast R}=-(K,\omega,v)_{-}^{\ast L};\quad 
k_{+}^{\ast R}=k_{-}^{\ast L}.
\end{equation}
They can be seen to be realized in Fig.~\ref{FIG.FRONTEN}(b) and 
Fig.~\ref{FIG.FRONTEN}(c).
 
The GLE approximation (\ref{K-GLE}) - (\ref{omega-GLE}) shows beyond the relations 
(\ref{symmet1})-(\ref{symmet3}) the following additional relations
\begin{equation}
k_{+}^{\ast T}=k_{-}^{\ast T}=k_c^T \quad ; 
\quad \omega_{+}^{\ast T}=\omega_{-}^{\ast T}=0
\end{equation}
\begin{equation} \label{GLE-Symmetrie2}
K_{+}^{\ast L}=-K_{-}^{\ast L}=K_{+}^{\ast R}=-K_{-}^{\ast R},
\end{equation}  
\begin{equation}
\omega_{+}^{\ast L}=\omega_{-}^{\ast L}=-\omega_{+}^{\ast R}=-\omega_{-}^{\ast R}
\end{equation}
that follow from the fact, that $\omega_c^T = v_g^T = c_0^T = c_1^T = 0$,
$k_c^R=k_c^L$, $v_g^R=-v_g^L$, 
$\tau_0^R=\tau_0^L$, $c_0^R=-c_0^L$, $c_1^R=-c_1^L$, and
$\xi_0^{R}=\xi_0^{L}$ for $Re=0$ \cite{PIN}.

 \section{Summary}
 We have determined the influence of an axial through-flow on the spatiotemporal
growth behavior of structurally different vortex perturbations of the basic 
Couette-Poiseuille flow in
the Taylor-Couette system with radius ratio $\eta=0.5$. To that end we
have solved the linearized NSE numerically with a shooting method for vortex 
perturbations with azmuthal wave numbers $m=0$ (TVF), $m=1$ (L-SPI), and $m=-1$
(R-SPI) in a wide range of the parameters $Re, R_1$, and $R_2$. Here symmetry
properties allowed us to restrict ourselves to positive through-flow Reynolds
numbers $Re$. For each of the three different vortex varieties we have
investigated ({\it i}) axially extended vortex structures with homogeneous 
amplitudes, ({\it ii}) axially localized vortex pulses consisting of a linear
superposition of axially extended vortex modes with different real axial wave
numbers $k$, and ({\it iii}) vortex fronts. 

Central to our analysis is the determination of the complex dispersion 
relations $\sigma(Q)$ of the linearized NSE for vortex modes with the three 
different $m$. We have evaluated $\sigma$ over the plane of complex wave 
numbers $Q =  k - i K$ for patterns ({\it ii}, {\it iii}) and along 
the real $k$-axis for pattern ({\it i}). We have also determined 
the Ginzburg-Landau amplitude equation approximation $\sigma_{GLE}(Q)$ in order
to analyze its predictions for the vortex stuctures ({\it ii}, {\it iii}) in 
comparison with the correct NSE dispersion relation $\sigma(Q)$. In each case 
symmetry relations are elucidated.

First we have
evaluated the critical bifurcation thresholds $R_{1c}(Re, R_2, m)$ for axially 
extended vortex structures. Then using a saddle-point analysis of $\sigma(Q)$
we have determined the boundaries $R_{1c-a}(Re, R_2, m)$ between absolute and
convective instability of the basic state at which one of the fronts of the
expanding vortex pulses reverts its propagation direction in the laboratory 
frame. Here we have elucidated also in some detail how the different saddle
topologies of $\sigma(Q)$ and of $\sigma_{GLE}(Q)$ explain some of the
shortcomings of the latter. Fit parameters for power-law expansions of
the reduced boundaries $\epsilon_c, \epsilon_{c-a}$, and $\epsilon_{c-a}^{GLE}$
up to $Re^4$ are listed in two tables. 

Finally we have determined the linearly selected front 
behavior of growing vortex patterns with $m=0, \pm 1$ for $R_2=0$ under two 
different types of front intensity envelopes: type $+$ shows growth in positive 
z-direction while type $-$ locates growth in negative z-direction. The 
combination of the
three different dynamics of the constituent vortex modes ($m=0, \pm 1$) and of
the two different spatial intensity profiles ($+$, $-$) leads to six different 
fronts $\sim e^{i (k^\ast z - \omega^\ast t)}e^{K^\ast (z - v^\ast t)}$
in the laboratory frame. Their velocity $v^\ast$, spatial growth rate $K^\ast$,
wave number $k^\ast$, and frequency $\omega^\ast$ as
determined via a saddle point analysis of the respective dispersion
relations differ in general from each other in the presence of a through-flow.

\acknowledgments
This work was supported by the Deutsche Forschungsgemeinschaft.

\clearpage
 
\clearpage

 \clearpage 
 
 \begin{table}[h] 
 \begin{center}
 \begin{tabular}{c|cccc} 
 &\multicolumn{4}{c}{Operation}\\
 &\ \ A&\ \ B&\ \ C&\ \ D \\
 \hline
 \hline
 $k \rightarrow$&$\quad k$&$\ -k$&$\ -k$&$\quad k$\\
 $m \rightarrow$&$\quad m$&$\ -m$&$\quad m$&$\ -m$\\
 $Re \rightarrow$&$\quad Re$&$\quad Re$&$\ -Re$&$\ -Re$\\
 $\sigma \rightarrow$&$\quad \sigma$&$\quad {\overline \sigma}$&$\quad \sigma$&$\quad {\overline \sigma}$\\
 $U \rightarrow$&$\quad U$&$\quad {\overline U}$&$\quad U$&$\quad {\overline U}$\\
 $V \rightarrow$&$\quad V$&$\quad {\overline V}$&$\quad V$&$\quad {\overline V}$\\
 $W \rightarrow$&$\quad W$&$\quad {\overline W}$&$\ -W$&$\ -{\overline W}$\\
 $X \rightarrow$&$\quad X$&$\quad {\overline X}$&$\quad X$&$\quad {\overline X}$\\
 $Y \rightarrow$&$\quad Y$&$\quad {\overline Y}$&$\quad Y$&$\quad {\overline Y}$\\
 $Z \rightarrow$&$\quad Z$&$\quad {\overline Z}$&$\ -Z$&$\ -{\overline Z}$\\
 \end{tabular}
 \end{center}
 \caption{Transformation behavior of the eigenvalues and eigenfunctions of the
 eigenvalue problem (\ref{resultEW} - \ref{defineL}) under symmetry operations.
 Here A denotes the identity, B complex conjugation, C axial reflection, and 
 D complex conjugation (indicated by an overbar) combined with axial reflection.
 \label{TAB.MOD-SYMM}}
 \end{table}
 \begin{table}[h] 
 \begin{tabular}{|c||c|c|c|c|c|c|c|c|c|}
 \hline
 $R_{2}$&-150&-125&-100&-75&-50&-25&0&25&50\\
 \hline
 \hline
 \multicolumn{10}{|c|}{$\epsilon_{c}=a_{2}Re^{2}+a_{4}Re^{4}$} \\
 \hline
 $a_{2}*10^{4}$&0.590&0.795&1.089&1.502&2.083&3.115&3.679&2.447&1.307\\
 \hline
 $a_{4}*10^{9}$&5.181&3.966&1.151&-2.804&-3.385&-15.62&-34.97&-14.55&-0.958\\
 \hline
 \hline  
 \multicolumn{10}{|c|}{$\epsilon_{c-a}=a_{2}Re^{2}+a_{4}Re^{4}$} \\
 \hline
 $a_{2}*10^{3}$&1.181&1.451&1.854&2.516&3.906&5.588&6.238&4.338&2.483\\
 \hline
 $a_{4}*10^{6}$&0.116&0.139&0.182&0.293&-2.583&-4.642&-5.716&-3.682&-1.740\\
 \hline 
 \multicolumn{10}{|c|}{$\epsilon_{c-a}^{GLE}=a_{2}Re^{2}+a_{4}Re^{4}$} \\
 \hline
 $a_{2}*10^{3}$&1.161&1.416&1.800&2.517&4.068&6.547&7.732&5.152&2.762\\
 \hline
 $a_{4}*10^{6}$&-0.164&-0.268&-0.481&-1.011&-2.106&-3.807&-4.900&-3.022&-1.425\\
 \hline 
 \end{tabular}
 \caption{Fitparameters for the $Re$-dependence of the stability boundaries of
 the basic state against growth of TVF $(m=0)$ perturbations for different
 $R_2$: reduced critical bifurcation threshold $\epsilon_c$ of axially extended 
 vortex patterns and boundary $\epsilon_{c-a}$ between convectively and 
 absolutely unstable parameter regime. Here $\epsilon_{c-a}^{GLE}$ is obtained
 from the GLE approximation (cf. Sec.~\ref{SEC.GLA1}). 
 \label{TAB.FITPARA}}
 \end{table}
 
 \begin{table}
 \begin{tabular}{|c||c|c|c|c|c|c|c|c|c|}
 \hline
 $R_{2}$&-150&-125&-100&-75&-50&-25&0&25&50\\
 \hline
 \hline
 \multicolumn{10}{|c|}{$\epsilon_{c}=a_{1}Re+a_{2}Re^{2}+a_{3}Re^{3}+a_{4}Re^{4}$} \\
 \hline
 $a_{1}*10^{3}$&-1.582&-2.365&-3.513&-4.887&-6.143&-5.833&-3.372&-1.402&-0.514\\
 \hline
 $a_{2}*10^{4}$&0.957&1.199&1.496&1.925&2.520&3.174&3.197&1.806&0.581\\
 \hline
 $a_{3}*10^{7}$&-9.181&-9.001&-8.485&-7.678&-6.741&-4.172&-3.832&-3.572&-3.685\\
 \hline
 $a_{4}*10^{9}$&1.018&-0.378&-0.180&-0.563&-3.673&-21.77&-36.15&-5.707&19.92\\
 \hline
 \hline
 \multicolumn{10}{|c|}{$\epsilon_{c-a}=a_{0}+a_{1}Re+a_{2}Re^{2}+a_{3}Re^{3}+a_{4}Re^{4}$} \\
 \hline
 $a_{0}*10^{3}$&2.358&5.497&3.775&-1.237&4.095&12.25&18.96&10.09&2.610 \\
 \hline
 $a_{1}*10^{3}$&-3.268&-3.578&-3.048&-0.449&0.926&1.159&1.012&1.217&1.455 \\
 \hline
 $a_{2}*10^{3}$&1.532&1.964&2.618&3.539&4.351&5.535&6.036&4.357&2.603 \\
 \hline
 $a_{3}*10^{6}$&8.408&11.87&14.97&9.992&6.487&4.919&3.680&1.367&-0.143 \\
 \hline
 $a_{4}*10^{6}$&-0.169&-0.368&-0.815&-1.977&-2.809&-4.068&-4.827&-3.203&-1.549 \\
 \hline
 \multicolumn{10}{|c|}{$\epsilon_{c-a}^{GLE}=a_0+a_1Re+a_2Re^2+a_3Re^3+a_4Re^4$} \\
 \hline
 $a_{0}*10^{3}$&1.101&-2.541&1.405&2.529&3.458&3.998&3.641&1.320&0.062\\
 \hline
 $a_{1}*10^{3}$&-4.117&0.755&-1.044&-0.318&0.831&2.063&1.495&1.377&1.625\\
 \hline
 $a_{2}*10^{3}$&2.030&2.792&2.820&3.435&4.437&6.166&7.265&4.981&2.723 \\
 \hline
 $a_{3}*10^{6}$&-8.845&-30.34&-7.121&4.780&10.58&8.229&-8.191&-12.58&-10.03 \\
 \hline
 $a_{4}*10^{6}$&-0.822&-2.713&-1.136&-1.004&-1.674&-2.923&-3.799&-2.234&-0.854 \\
 \hline
 \end{tabular}
 \caption{Fitparameters for the $Re$-dependence of the stability boundaries of
 the basic state against growth of L-SPI $(m=1)$ perturbations for different
 $R_2$: reduced critical bifurcation threshold $\epsilon_c$ of axially extended 
 vortex patterns and boundary $\epsilon_{c-a}$ between convectively and 
 absolutely unstable parameter regime. Here $\epsilon_{c-a}^{GLE}$ is obtained
 from the GLE approximation (cf. Sec.~\ref{SEC.GLA1}). The results for 
 R-SPI $(m=-1)$ perturbations are obtained according to
Sec.~\ref{SEC.SYMM-EIGENVALUE} from those for L-SPI $(m=1)$ by 
$Re \rightarrow -Re$, i.e., by changing the sign of the odd coefficients in 
the table.
\label{TAB.FITPARA2}}
 \end{table}

\begin{thebibliography}{999}
\bibitem{epsT}
In the literature one uses also $\epsilon_T = \frac{T_{1}}{T_{1c}(Re=0)}-1$ as
control parameter where the inner cylinder's Taylor number $T_1$ is proportional
to $\Omega_1^2$. Thus, $\epsilon_T = 2 \epsilon + \epsilon^2$.
\end{thebibliography}
\end{document}